# Ultralow noise up-conversion detector and spectrometer at telecom band

Guo-Liang Shentu[1], Jason S. Pelc[2], Xiao-Dong Wang[3], Qi-Chao Sun[1], Ming-Yang Zheng[1], M. M. Fejer[2], Qiang Zhang[1], Jian-Wei Pan[1]

*1. Shanghai Branch, National Laboratory for Physical Sciences at Microscale and Department of Modern Physics,*
*University of Science and Technology of China, Shanghai, 201315, China*
*2. Edward L. Ginzton Laboratory, Stanford University, Stanford, California 94305*
*3. College of Physics and Electronic Engineering of Northwest Normal University,*
*Lanzhou,730070 China*
*qinagzh@ustc.edu.cn*

**Abstract:** We demonstrate upconversion single-photon detection for the 1550-nm band using a PPLN waveguide, long-wavelength pump, and narrowband filtering using a volume Bragg grating. We achieve total-system detection efficiency of around 30% with noise at the dark-count level of a silicon APD. Based on the new detector, a single-pixel up-conversion infrared spectrometer with a noise equivalent power of -142 dBm was demonstrated, which was better than liquid nitrogen cooled InGaAs array.

---

**1. Introduction**

Single-photon detectors (SPD) operating in the 1.55-μm telecom band are a core part of quantum communication over optical fiber [1]. Currently, SPD performance limits the transmission distance and key rate of fiber based quantum key distribution (QKD) [2,3].

Current telecom band commercial SPDs based on InGaAs/InP avalanche photodiodes (APDs) suffer from low efficiency about 10% and large dark-count rate of $10^3$ cps [4]. In contrast, the efficiency of Silicon APD working in the visible to NIR windows is up to 65% and the dark count rate is can be lower than 100 cps. Upconversion detectors [5-9] utilizing sum-frequency generation (SFG) in periodically poled lithium niobate (PPLN) waveguide or bulk crystals can convert the telecom-band photons into the near infrared window and then detect them with a Silicon APD. The detection efficiency of PPLN waveguide-based up-conversion detectors can be higher than 40% [7]. However, for early implementations pumped at short pump wavelengths, the noise count rate of the detector reached $10^5$ cps [7, 8]. Monochromator based filtering has been exploited to decrease the noise. However, due to the low throughput of the monochromator, the detection efficiency also reduced dramatically [10].

Noise in upconversion detectors mainly comes from spontaneous parametric down conversion (SPDC) and spontaneous Raman scattering (SRS) when the strong pump goes through the waveguide. A recent experiment demonstrated that a long wave pump can suppress both spontaneous processes [11,12] without dramatically reducing the detection efficiency. Here, we combine the long wave pump and a high efficiency narrow-band filter, volume Bragg grating (VBG) [13].

Meanwhile, based on upconversion detector, upconversion spectrometer has been invented to measure the infrared spectrum, whereby the pump wavelength is tuned and the phase-matching acceptance bandwidth of the upconverter acts as a frequency selective element [14]. The spectrometer had a noise equivalent power (NEP) of -128 dBm [15] and found an immediate application in testing quantum dot with a wavelength of 1.3 µm [16]. Based on our up-conversion detector, we improved the NEP by one order of magnitude due to noise reduction, i.e. -142 dBm, which is as good as the liquid nitrogen cooled InGaAs array based spectrometer [17]. Furthermore, the narrow band VBG helps to improve our spectrometer's resolution into 0.16 nm.

## 2. Upconversion detector

We fabricated PPLN waveguides via the reverse proton exchange technique. The waveguides are 52-mm long and are poled with a quasi-phase-matching (QPM) period of 19.6μm. The waveguides incorporated mode filters designed to match the mode size of SMF-28 optical fiber and were fiber pigtailed with coupling losses of approximately 0.7 dB, and had a total fiber-to-output-facet throughput of−1.5 dB. The waveguides were antireflection coated to avoid interference fringes and improve the system throughput.

A schematic of our experimental setup is shown in Fig. 1. We used a single longitudinal mode Thulium fiber laser (TDFL) and amplifier (TDFA) manufactured by AdValue Photonics (Tucson, AZ) as the pump source for this experiment. The laser amplifier emits an average power of approximately 800 mW with a wavelength of 1950 nm. The pump-laser power was controlled by adjusting the pump current of TDFA, and noise emission from the pump laser was blocked using a 1.55-µm/1.9-µm wavelength division multiplexer (WDM).

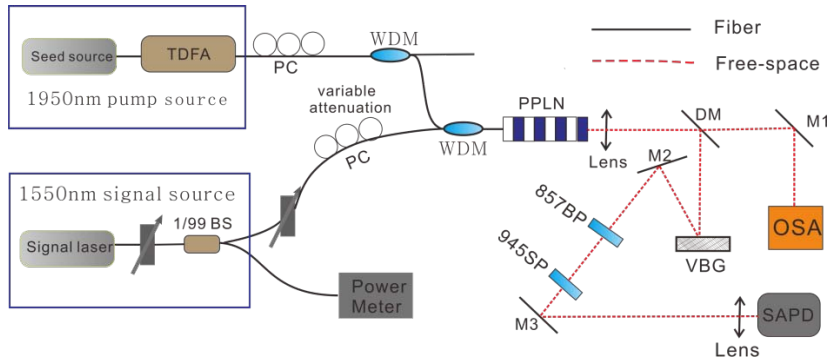

Fig. 1. Schematic of the noise-free up-conversion single photon detector. BS: beam splitter, M1-M3: mirrors.

A cw, single-frequency, tunable telecom-band external cavity diode laser (ECDL) provided the signal at a wavelength of 1.55 μm. We used a variable attenuator, a 1/99 tap coupler and a calibrated detector to monitor the input signal power. The signal was combined with the pump in another 1.55-µm/1.9-µm WDM and coupled into the PPLN waveguide through the fiber pigtail. The waveguide only supports TM-polarized light so the polarizations of both the pump and the signal are controlled by polarization controllers (PCs) respectively before entering the waveguide. A Peltier cooler based temperature-control system is used to keep the waveguide's temperature at 56℃ to maintain the phase-matching condition.

The generated SFG photons are collected by an AR-coated objective lens, and are separated from the pump by a dichroic mirror (DM). The residual pump is transmitted through the DM and analyzed by an optical spectrum analyzer (OSA). A 945nm short pass filter (SPF) and a 857nm band pass filter (BPF) are used to block the second and higher order harmonic of the pump. A VBG with a 95% reflection efficiency is exploited to further suppress the noise from SRS. In our setup, VBG filter's spectral bandwidth is 0.05 nm (full width at half maximum), ten times narrower than our waveguide's acceptance bandwidth, which helps to reduce the SRS noise. Finally, the SFG photons are collected and detected by a Silicon APD (SAPD) with dark count of 25 cps.

The experimental result is shown in Fig. 2. We tune the pump power by adjusting the variable attenuator and record the detection efficiency and noise count rates. When the pump power is set at 58mW [18], the detection efficiency was 28.6%, with a noise count rate of 100 cps. When the pump power is set at 20 mw, the noise reduces to 25 cps, while the detection efficiency is still 15%.

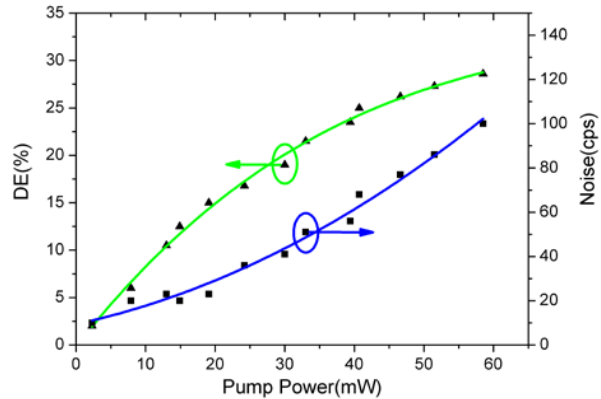

Fig. 2. The detection efficiency (green line, triangle) and noise count rate (blue line, square) versus pump power, which is measured just after the waveguide.

## 3. Spectrometer

Previous research utilized the PPLN waveguide's acceptance bandwidth as a filter in the frequency domain [14,15]. Scanning the pump wavelength, the center wavelength of this filter, allows us to trace out the spectrum of the signal light. In order to realize the up conversion spectrometer, we slightly changed the setup in Fig. 1 by replacing the 1950 nm seed fiber laser with a 1.9-μm band tunable ECDL. By adjusting the grating inside the 1.9-μm band ECDL, the emission spectrum of pump could be scanned from 1920 nm to 1980 nm, and thus the spectral range of the up conversion spectrometer is from 1532.9 nm to 1570.9 nm.

One of the advantages of our spectrometer is its low NEP ($NEP = \hbar\upsilon\sqrt{D}/\eta$). Here, η represents detection efficiency and D represents dark count rate. In our experiment, we set the pump power at 30 mW and the detection efficiency, dark count rate and NEP are 20%, 60 Hz and -142 dBm, respectivley.

In previous experiments, the resolution of the spectrometer is determined by the spectral acceptance of the PPLN waveguide [14,15]. In our setup, VBG filter's spectral bandwidth is 0.05 nm, ten times narrower than our waveguide's acceptance bandwidth. Therefore, our upconversion spectrometer's resolution is enhanced to 0.16nm. A point to note here is that the upconverted wavelength will change with different signal wavelength. For measurements of signals over total bandwidth greater than 3.09 nm, the angle of the VBG must be adjusted such that its reflection resonance tracks the changing upconverted wavelength. This can be achieved by mounting the VBG on a motorized rotation state.

In order to test our up conversion spectrometer, we used the light from a 1.55-μm multimode laser diode as the signal for the spectral measurements. The experimental results are shown in Fig 3a. We first took a spectral measurement (shown in black solid line) of the 1.5-μm laser diode using the commercial OSA when the output power of the laser diode was -15dBm. Before injecting the signal into the up-conversion spectrometer, we set the input signal power to -98.9 dBm, corresponding to approximately 1 million photons per second, by adjusting the tunable attenuator. We set the pump power at 30 mW and scan the pump laser and achieve a raw spectrum as shown in Figure 3(red dot line) [15]. Then we achieved the real spectrum (green dotted line) by a deconvolution with VBG's spectral curve (shown in the Figure 3b) [15].

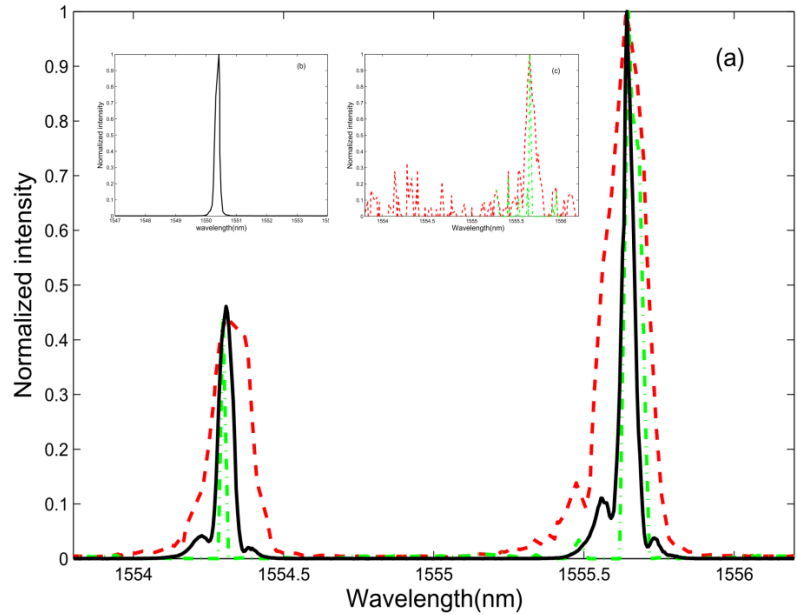

Fig. 3. Spectrum detected by the OSA (black solid line), raw spectrum (red dot line), deconvolved spectrum (green dotted line) by upconversion spectrometer. (a) Input signal power of -98 dBm. (b) Reflection spectrum from the VBG. (c) Input signal power of -135 dBm.

In order to verify the NEP of our upconversion spectrometer, the peak of the spectrum was still measured well down to an input power of -135 dBm and the spectrum was Fig. 3c.

## 4. Conclusion

We utilizing long-wavelength pump and VBG filter to achieve a high efficiency and ultralow noise up-conversion single photon detector. The new detector can be directly used in a QKD system to improve its performance. Based on the detector, we demonstrate an upconversion spectrometer with a NEP of -142 dBm and a resolution of 0.16 nm. Moreover, the ultralow noise frequency conversion technique can find other important applications, for example, erasing the frequency difference between two photons to enable interference measurements [19] and bridging the quantum memory with the telecom band wavelength [20-24].

**Acknowledgement**

This work has been supported by the National Basic Research Program of China (under Grant No. 2013CB336800, 2011CB921300 and 2011CBA00300), the NNSF of China, the CAS, and the Shandong Institute of Quantum Science & Technology Co., Ltd. J.S.P. and M.M.F. acknowledge the U.S. AFOSR for their support under Grant No. FA9550-09-1-0233.